\documentclass[%
reprint,
nofootinbib,
 amsmath,amssymb,
 aps,
]
{revtex4-1}
\usepackage{epsfig,epsf,rotating,wrapfig,tabularx}
\usepackage{dcolumn}
\usepackage{bm}
\usepackage{graphicx}
\usepackage{color}
\usepackage{subfigure}

\begin{document}

\preprint{APS/123-QED}

\title{Two components in charged particle production in heavy-ion collisions.}

\author{\firstname{A.~A.}~\surname{Bylinkin}}
 \email{alexandr.bylinkin@cern.ch}
\affiliation{%
Moscow Institute of Physics and Technology, MIPT, Moscow, Russia\\
National Research Nuclear University MEPhI, Moscow, Russia
}%
\author{\firstname{N.~S.}~\surname{Chernyavskaya}}
 \email{nadezda.chernyavskaya@cern.ch}
\affiliation{%
Moscow Institute of Physics and Technology, MIPT, Moscow, Russia\\
 Institute for Theoretical and Experimental
Physics, ITEP, Moscow, Russia\\
National Research Nuclear University MEPhI, Moscow, Russia
}%
\author{\firstname{A.~A.}~\surname{Rostovtsev}}
 \email{rostov@itep.ru}
\affiliation{%
 Institute for Information Transmission Problems, IITP, Moscow, Russia
}%


\begin{abstract}
Transverse momentum spectra of charged particle production in heavy-ion collisions are considered in terms of a recently introduced Two Component parameterization combining exponential (``soft'') and power-law (``hard'') functional forms. The charged hadron densities calculated separately for them are plotted versus number of participating nucleons, $N_{part}$. The obtained dependences are discussed and the possible link between the two component parameterization introduced by the authors and the two component model historically used for the case of heavy-ion collisions is established. Next, the variations of the parameters of the introduced approach with the center of mass energy and centrality are studied using the available data from RHIC and LHC experiments. The spectra shapes are found to show universal dependences on $N_{part}$ for all investigated collision energies.
 
\end{abstract}

\pacs{Valid PACS appear here}
\maketitle

\section{Introduction}
Two-component models have been used in heavy-ion phenomenology for a long time. The reason for that is that there is no single theoretical approach that can simultaneously describe both low-$p_T$ and high-$p_T$ hadron production. The main object of study of such models~\cite{WG} is charged particle density, $dN_{ch}/d\eta$, which is expected to  scale with number of participating nucleons, $N_{part}$, or number of binary parton-parton collisions, $N_{coll}$, for ``soft'' and ``hard'' regimes of particle production, respectively. Such scaling becomes a subject of various phenomenological discussions - linear scaling with $N_{part}$ is expected for ``soft'' processes, while scaling with $N_{coll}$ is expected for the ``hard'' regime of particle production~\cite{ALICE1}\footnote{Also note that $N_{coll} \propto N_{part}^{4/3}$.}:
\begin{equation}
\label{eq:tc}
dN_{ch}/d\eta = A\cdot N_{part} + B\cdot N_{coll}.
\end{equation}

Recently, another two component approach accounting for another aspect of charged particle production - transverse momentum spectra $d^2\sigma/dp_T^2d\eta$ -  has been introduced by the authors ~\cite{OUR1}. Remarkably, it was also suggested to consider two sources of hadroproduction related to ``soft'' and ``hard'' regimes, respectively, and therefore parameterize transverse momentum spectra by a sum of an exponential (Boltzmann-like) and a power-law $p_T$ distributions:
\begin{equation}
\label{eq:exppl}
\frac{\mathrm{d}^2\sigma}{\mathrm{d} \eta \mathrm{d}p_{T}^2} = A_e\exp {(-E_{Tkin}/T_e)} +
\frac{A}{(1+\frac{p_T^2}{T^{2}\cdot N})^N},
\end{equation}
where $E_{Tkin} = \sqrt{p_T^2 + M^2} - M$
with M equal to the produced hadron mass and $A_e, A, T_e, T, N$ are the free parameters to be determined by fit to the data.

Moreover, this approach was shown to effectively describe heavy-ion collision data ~\cite{OURI} when the exponential term of (\ref{eq:exppl}) is substituted with the well-known Blast-Wave formula~\cite{Hydro}:
\begin{equation}
\label{eq:Bessel}
\begin{split}
\frac{\mathrm{d n}}{\mathrm{d} \eta \mathrm{d}p_{T}^2} \propto 
\int_0^R r \text{ } \mathrm{d}r \text{ }m_{T} \text{ } I_{0}\left( \displaystyle \frac{p_{T} \sinh \rho  }{T_e} \right) K_{1} \left( \frac{ m_{T}   \cosh \rho }{T_e} \right) \\ ,
\end{split}
\end{equation}
taking into account hydrodynamical expansion of the colliding system. In this approach the expanding under the pressure in the longitudinal direction system generates the transverse flow. The particle distribution is considered to be Boltzmann again but in the local fluid rest frame. In (\ref{eq:Bessel}) $\rho = tanh^{-1}\beta_r$ and $\beta_r(r) = \beta_s(\frac{r}{R})$, with $\beta_s$ standing for the surface velocity, $m_T = \sqrt{m^2 + p_T^2}$, $I_0$ and $K_1$ are the modified Bessel functions. 

 In~\cite{OURI} it was also shown that an additional power-law term is needed to describe the charged hadron spectra in central PbPb collisions in the full range of transverse momenta. Thus, the experimental data are fitted to the function:
 \begin{equation}
\label{eq:Besselplpl}
\begin{split}
\frac{\mathrm{d^2}\sigma}{\mathrm{d} \eta \mathrm{d}p_{T}^2} = A_e \cdot 
\int_0^R r \text{ } \mathrm{d}r \text{ }m_{T} \text{ } I_{0}\left( \displaystyle \frac{p_{T} \sinh \rho  }{T_e} \right) K_{1} \left( \frac{ m_{T}   \cosh \rho }{T_e} \right)\\
 + \frac{A}{(1+\frac{p_T^2}{T^2\cdot N})^N} 
 +\frac{A_{1}}{(1+\frac{p_T^2}{T^2_1\cdot N_1})^{N_1} }.
 \end{split}
\end{equation}

\begin{figure}[!ht]
\includegraphics[width =8.5cm]{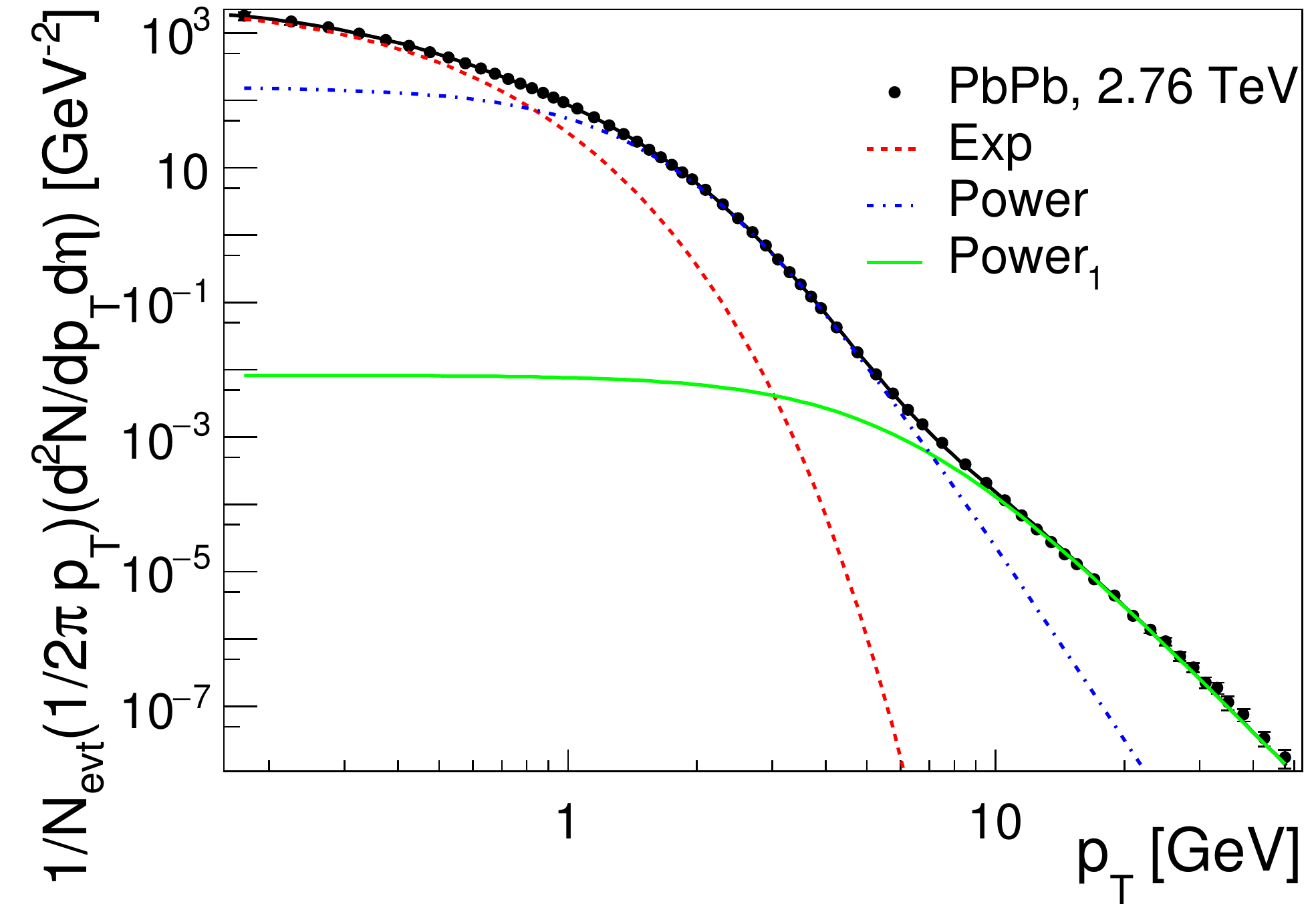}
\caption{\label{fig.0} (Color online) Central lead-lead collisions~\cite{ALICE}: the red (dashed) line shows the hydrodynamic term~(\ref{eq:Bessel}) and the green (solid) and blue (dash-dot) lines - two power-law terms (\ref{eq:Besselplpl}).}
\end{figure}

A typical charged particle spectrum as a function of transverse momentum measured by the ALICE collaboration in PbPb collisions is shown in figure~\ref{fig.0} together with the introduced new three component fit function. 

While the high$-p_T$ tail  (shown with the green line in figure~\ref{fig.0}) has the same slope parameter $N$ as in $pp-$collisions~\cite{OURI}\footnote{This observation is also supported by the fact that the nuclear modification factor $R_{AA}$ comes to a constant value for the high-$p_T$ region.} and can be rather well reproduced in Monte Carlo generators, the  low-$p_T$(``soft'' part) and mid-$p_T$(``hard'') parts  (shown with red and blue lines in figure~\ref{fig.0})  of the spectra have only phenomenological descriptions.
Therefore, in this paper we suggest to look how these two contributions and their parameter values vary with centrality and center of mass energy to better understand the dynamics of charged hadron production in heavy-ion collisions.
 
\section{Charged hadron densities}
\label{density}
First of all, recall that integrating the transverse momentum spectrum $d^2\sigma/dp_T^2d\eta$ by $dp_T^2$ from $0$ to the upper kinematic limit, one gets the charged particle densities $d\sigma/d\eta$ \footnote{The $\eta_{lab} = 0$ region is considered.}. Therefore, the possible link between the discussed two component approaches (\ref{eq:tc}) and (\ref{eq:exppl}) can be established giving the further insight into the interplay of ``soft'' and ``hard'' regimes of hadroproduction in heavy-ion collisions.

\begin{figure}[!ht]
\includegraphics[width =8.5cm]{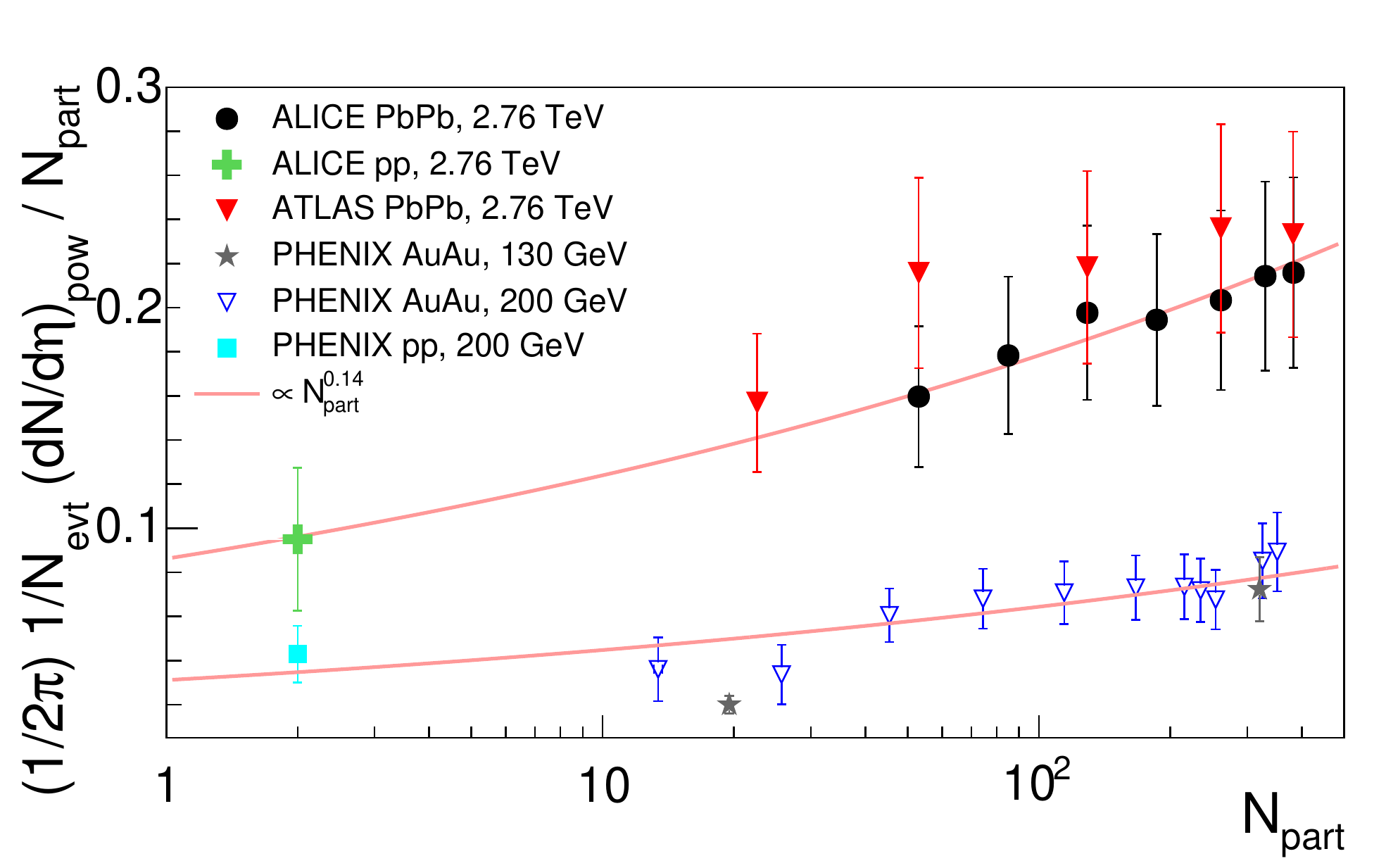}
\caption{\label{fig.3pow} (Color online) The charged particle densities divided over the number of participating nucleons $N_{part}$ obtained for the mid-$p_T$ part (blue line in the figure~\ref{fig.0}) of the spectra as a function of $N_{part}$. The lines show the scaling of this quantity as $N_{part}^{0.14}$ the same for RHIC~\cite{PHENIX} and LHC~\cite{ALICE} data.}
\end{figure}

\begin{figure}[!ht]
\includegraphics[width =8.5cm]{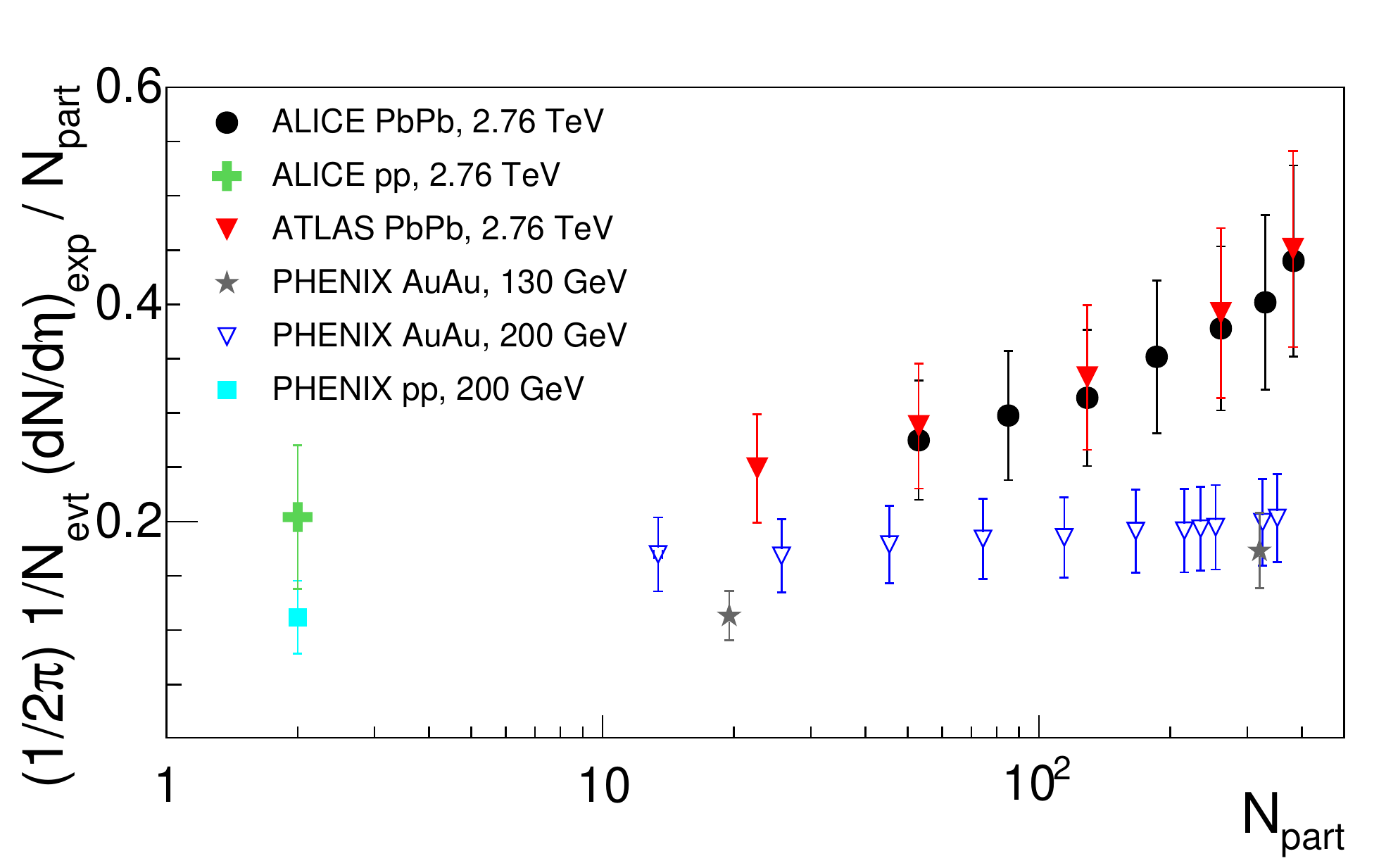}
\caption{\label{fig.3exp} (Color online) The charged particle densities divided over the number of participating nucleons $N_{part}$ obtained for the ``thermal'' part (red line in the figure~\ref{fig.0}) of the spectra as a function of $N_{part}$.}
\end{figure}

Let us now have a look how charged hadron densities $dN_{ch}/d\eta$ vary with the centrality of heavy-ion collisions.  Therefore, to study this scaling we suggest to plot charged hadron densities $1/N_{evt} dN /d\eta / N_{part}$ calculated  separately for ``soft'' and ``hard'' parts of (\ref{eq:Besselplpl}) as functions of $N_{part}$ as shown in the figures~\ref{fig.3pow} and \ref{fig.3exp}.

As one can notice, in figure \ref{fig.3pow} charged hadron densities obtained for the power-law term contribution show an universal scaling with $N_{part}^{\alpha} (\alpha = 1.14)$  independent on the collision energy, with $\alpha > 1$ indicating that the mid-$p_T$ power-law terms of (\ref{eq:exppl}), (\ref{eq:Besselplpl}) are related to some ``hard" regime as postulated in our model. The value of $\alpha < 1.33$ which might be expected for the purely ``hard'' regime of hadroproduction\footnote{$dN_{ch}/d\eta\propto N_{part}^{1.33}\propto N_{coll}^{1.0}$} is explained by the quenching of mini-jets in heavy-ion collisions.  The rescattering of the original jet diminishes the mean $p_T$ of the jet but enlarges the number of the final mini-jets, that leads to a larger particle density $dN/d\eta$. This is the origin of an additional (in comparison with (\ref{eq:exppl})) power-law term in (\ref{eq:Besselplpl}) which contribution is shown by the blue line in figure~\ref{fig.0}.

At first sight the exponential part caused in $pp$-case just by the incoming valence quarks~\cite{OURQ} should be proportional to $N_{part}$. This is more or less consistent with the behavior observed in figure \ref{fig.3exp} for the RHIC energies. However, at a higher, LHC energy, the energy density of the partons increases and the number of final state rescatterings becomes so large that the secondaries produced in these rescatterings start to thermalize. In our fit this contribution is described by the first term of (\ref{eq:Besselplpl}). By this reason the 'thermal' part of the particle density $dN/d\eta$ increases with $N_{part}$, especially at the LHC energy.

\begin{figure}[!ht]
\includegraphics[width =8.5cm]{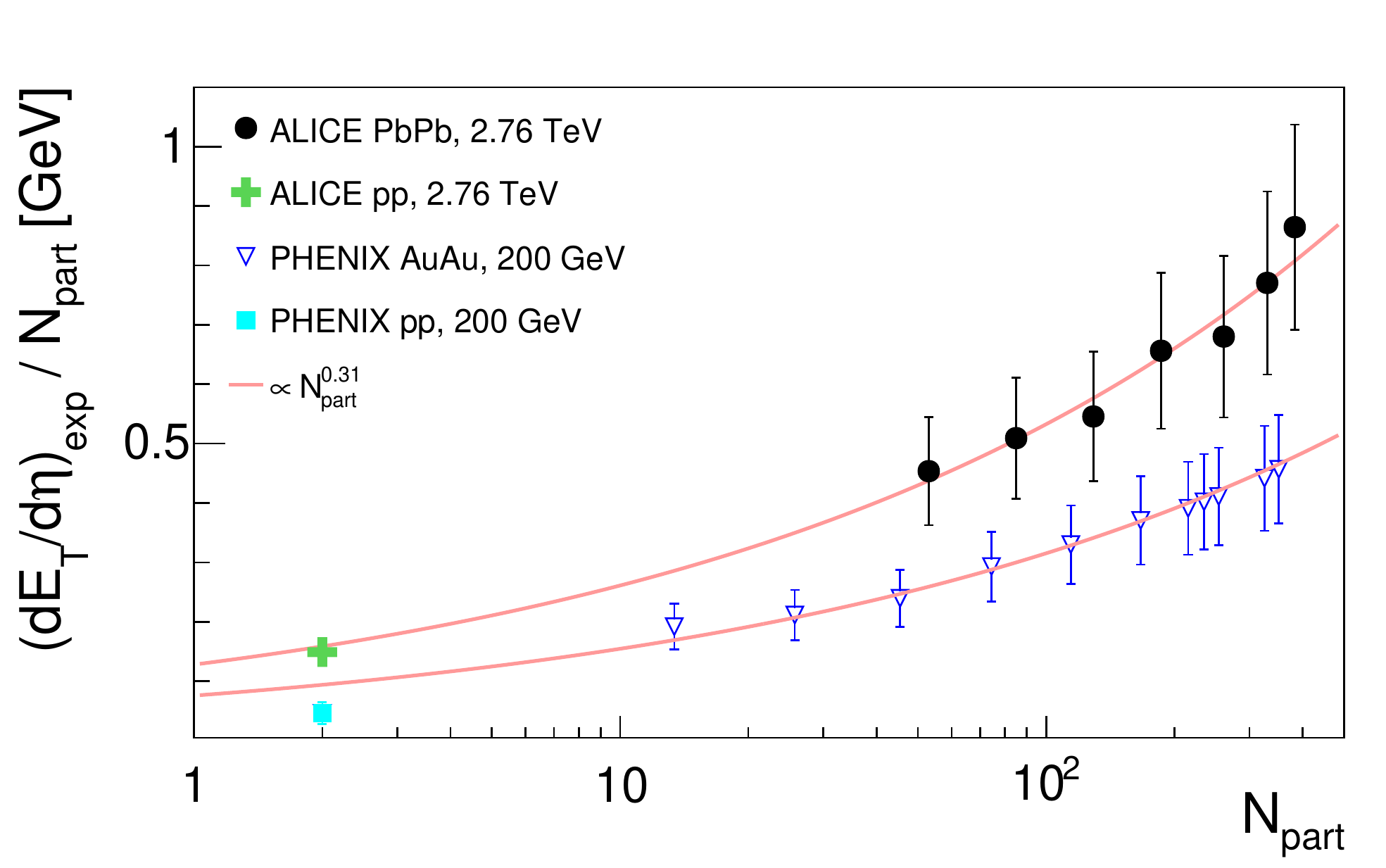}
\caption{\label{fig.4} (Color online) The total transverse energy for ``soft'' particle production (red line in the figure~\ref{fig.0})  divided over the number of participating nucleons $N_{part}$ as a function of $N_{part}$.  The lines show the scaling of this quantity as $N_{part}^{0.31}$ the same for RHIC and LHC data.}
\end{figure}

To perform the quantitative test we suggest to calculate the total transverse energy $\Sigma dE_T/d\eta$ for the ``thermal'' part of the spectra. This quantity divided over the number of participants is shown in figure~\ref{fig.4}. One can see that the total transverse energy $\Sigma dE_T/d\eta$ has the same scaling with $N_{part}^{1.31}$ for both RHIC and LHC data which confirms the picture described above.

Remarkably, the observed scaling  $\Sigma dE_T/d\eta\propto N_{part}^{1.31} \approx N_{coll}^{1.0}$ correlates with the fact that naively $\Sigma dE_T/d\eta$ should be proportional to the total energy stored in the colliding system. This observation further supports the idea that the two component model indeed reveal  the underlying hadroproduction dynamics in heavy-ion collisions and not just a useful ansatz as was argued in \cite{PHENIX2}.

\section{Parameters}
\label{parameters}
Let us now have a look how the values of the parameters $T$ and $N$ of the power-law term of (\ref{eq:Besselplpl}) (blue line in figure~\ref{fig.0}) depend on the energy and the centrality of a collisions.
In~\cite{OURI} it was suggested that the mini-jets travelling through the dense medium should loose their energy because of the multiple rescatterings. This effect, in turn, will result in higher values of $T$ and $N$ parameters (of the mid-$p_T$ term), as expected for higher thermalization. Here we consider $pp-$collisions as a reference point for the case of heavy-ion collisions to study the modification of the spectra with centrality. On the other hand, the most universal parameter characterizing the centrality of a heavy-ion collision is $N_{part}$ or the number of participating nucleons related to the impact parameter. 
 Therefore, we propose to look at the variation of the ratios $T/T_{pp}$ and $N/N_{pp}$ obtained from the fit to the experimental data as a function of $N_{part}$, where $T_{pp}$ and $N_{pp}$ stand for the parameter values obtained for $pp-$collisions at the same c.m.s. energy per nucleon $\sqrt{s}/N$. 

\begin{figure}[!ht]
\includegraphics[width =8.5cm]{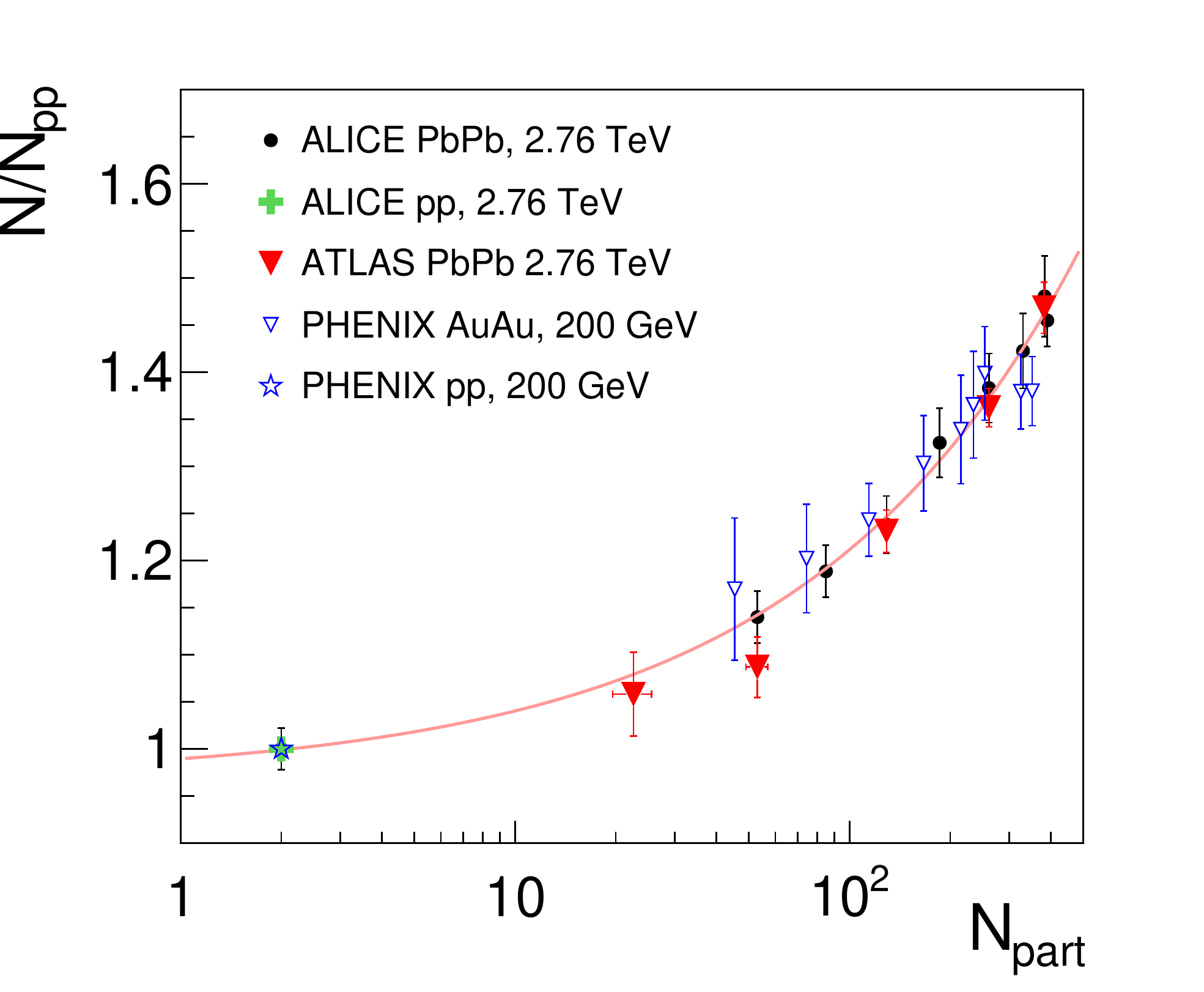}
\caption{\label{fig.1} (Color online) The dependence of the ratio of the parameters $N$ and $N_{pp}$ of the two-component model on $N_{part}$. The $N$ is obtained for heavy-ion collisions and $N_{pp}$ for pp-collisions at the same c.m.s. energy. The red (solid) line shows the dependence~(\ref{eq:NNcolll}).}
\end{figure}

\begin{figure}[!ht]
\includegraphics[width =8.5cm]{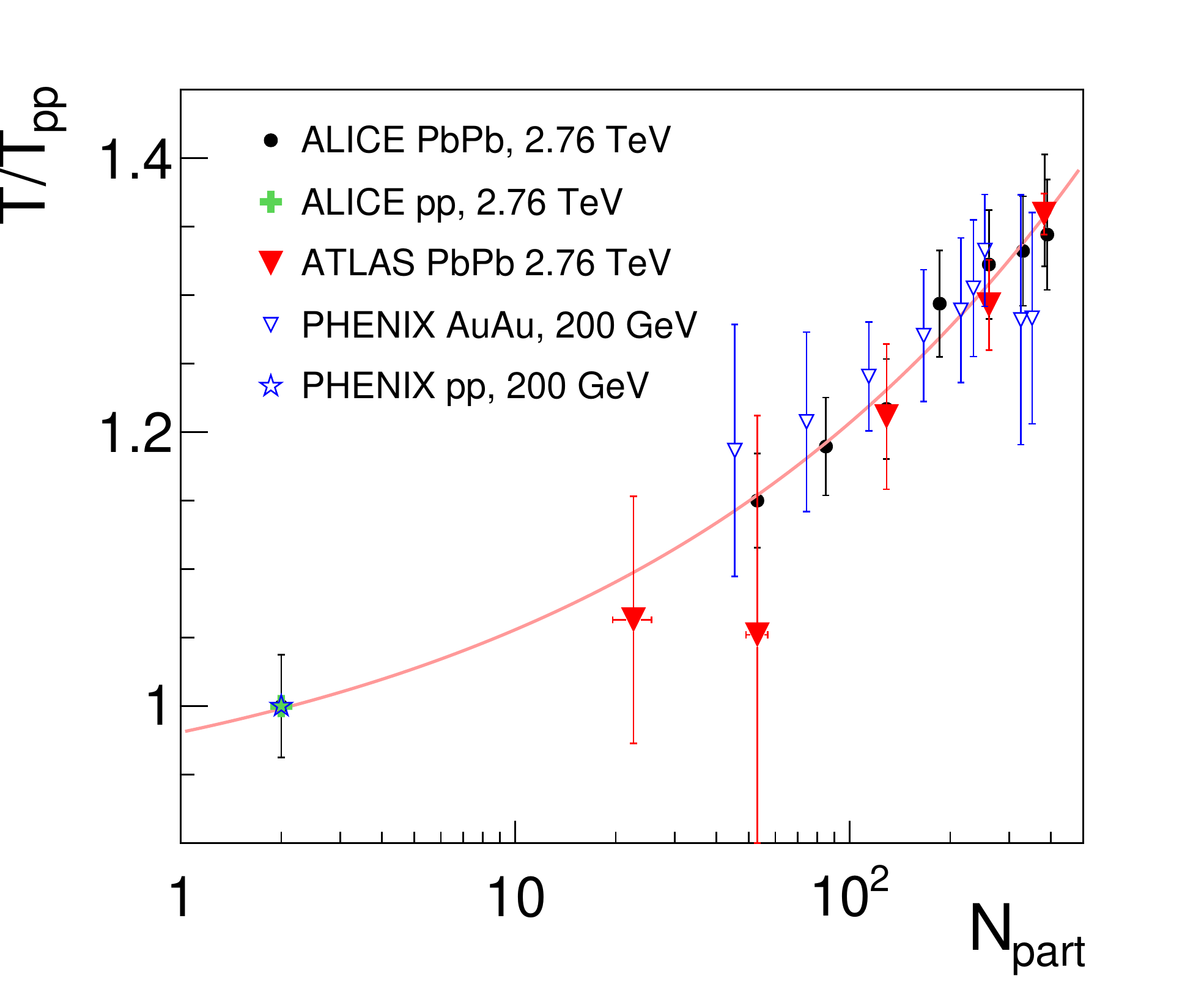}
\caption{\label{fig.2} (Color online) The dependence of the ratio of the parameters $T$ and $T_{pp}$ of the two-component model on $N_{part}$. The $T$ is obtained for heavy-ion collisions and $T_{pp}$ for pp-collisions at the same c.m.s. energy. The red (solid) line shows the dependence~(\ref{eq:TNcolll}). }
\end{figure}
 
 As shown in figures~\ref{fig.1}-\ref{fig.2}, indeed, both of the parameters, $T$ and $N$, grow with $N_{part}$ as it was natural from our naive expectations. Using the data from the ALICE~\cite{ALICE}, ATLAS~\cite{ATLAS} and PHENIX~\cite{PHENIX} experiments their behavior can be parameterized in the following way as a function of $N_{part}$:
 \begin{equation}
\label{eq:NNcolll} 
N / N_{pp} = 0.02 \cdot N_{part}^{0.53} + 0.97,
\end{equation}
 \begin{equation}
\label{eq:TNcolll}
T / T_{pp} = 0.07 \cdot  N_{part}^{0.30} + 0.91.
\end{equation}
Remarkably, in figures~\ref{fig.1}-\ref{fig.2} the points corresponding to $pp$-collisions are nicely placed on these trends (\ref{eq:NNcolll})-(\ref{eq:TNcolll}) extracted from the heavy-ion data. Moreover, from these figures~\ref{fig.1}-\ref{fig.2} one can notice a universal scaling of the parameters $T$ and $N$ with $N_{part}$ to be the same both for RHIC and LHC data.  This supports the suggested picture of hadroproduction in heavy-ion collisions and therefore one can conclude that the change of the spectra shape with centrality is indeed related to the higher suppression of the mini-jets produced in a collision. 
\vspace{1cm} 

\section{Conclusion}
In conclusion, transverse momentum spectra in heavy ion collisions have been considered using the two component parameterization. Scaling of exponential and power-law contributions to the charged hadron spectra with centrality in heavy-ion collisions were studied separately. The charged hadron densities for ``hard'' part turned out to have the same scaling for RHIC and LHC data with $\alpha > 1$ indicating the ``hard'' regime of hadroproduction. For the ``soft'' part the deviation from $\alpha \approx 1$ was found, especially for the LHC, and is explained by large initial parton densities which strongly enhance the rescattering of the mini-jets leading to the thermalization of the part of secondaries, that is to the additional contribution to the ``soft'' term. Finally, the variations of the parameters obtained from the fit have been studied as a function of $N_{part}$ and c.m.s. energy. The same universal dependence of the parameters of the model, $T$ and $N$,  for RHIC and LHC data was found giving further insight into the hadroproduction dynamics in heavy-ion collisions.

\begin{acknowledgements}
The authors thank Professor Mikhail Ryskin for fruitful discussions and his help provided during the preparation of this paper. A. Bylinkin acknowledges support from the Ministry of Education and Science of the Russian Federation (Contract No. 02.А03.21.0003 dated of August 28, 2013). A. Rostovtsev thanks the Russian Foundation for Sciences for financial support provided for this research (project № 14-50-00150). 
\end{acknowledgements}


\begin{thebibliography}{99}
\bibitem{WG}
  X.~N.~Wang and M.~Gyulassy,
  Phys.\ Rev.\ Lett.\  {\bf 86} (2001) 3496
  [nucl-th/0008014].

\bibitem{ALICE1}
  B.~Abelev {\it et al.}  [ALICE Collaboration],
  Phys.\ Rev.\ C {\bf 88} (2013) 4,  044909
  [arXiv:1301.4361 [nucl-ex]].

\bibitem{OUR1}
  A.~A.~Bylinkin and A.~A.~Rostovtsev,
  Phys.\ Atom.\ Nucl.\  {\bf 75} (2012) 999
   Yad.\ Fiz.\  {\bf 75} (2012) 1060;\\
A.~A.~Bylinkin and A.~A.~Rostovtsev,
  arXiv:1008.0332 [hep-ph].


\bibitem{OURI}
  A.~A.~Bylinkin, N.~S.~Chernyavskaya and A.~A.~Rostovtsev,
  Phys.\ Rev.\ C {\bf 90} (2014) 1,  018201
  [arXiv:1405.3055 [hep-ph]].


\bibitem{Hydro}
  E.~Schnedermann, J.~Sollfrank and U.~W.~Heinz,
  Phys.\ Rev.\ C {\bf 48} (1993) 2462
  [nucl-th/9307020].



\bibitem{ALICE}
  B.~Abelev {\it et al.}  [ALICE Collaboration],
  Phys.\ Lett.\ B {\bf 720} (2013) 52
  [arXiv:1208.2711 [hep-ex]].


\bibitem{PHENIX}
 S.~S.~Adler {\it et al.}  [PHENIX Collaboration],
  Phys.\ Rev.\ C {\bf 69} (2004) 034910
  [nucl-ex/0308006].

  
\bibitem{OURQ}
  A.~A.~Bylinkin and A.~A.~Rostovtsev,
  ``Role of quarks in hadroproduction in high energy collisions,''
  Nucl.\ Phys.\ B {\bf 888} (2014) 65
  [arXiv:1404.7302 [hep-ph]].

\bibitem{PHENIX2}
  S.~S.~Adler {\it et al.} [PHENIX Collaboration],
  Phys.\ Rev.\ C {\bf 89} (2014) 4,  044905
  [arXiv:1312.6676 [nucl-ex]].


\bibitem{ATLAS}
  G.~Aad {\it et al.}  [ATLAS Collaboration],
  arXiv:1504.04337 [hep-ex].
  



\end{thebibliography}
\end{document}